\documentclass{article}
\usepackage[a4paper, total={6in, 8in}]{geometry}
\usepackage{xcolor}
\usepackage{amsmath}
\usepackage{hyperref}
\usepackage{authblk}
\usepackage{blindtext}
\usepackage{graphicx}
\hypersetup{
    colorlinks=true,
    citecolor=red,
    linkcolor=blue,
    }

\usepackage[backend=biber, style=nature, backref=true]{biblatex}

\title{Biophysical and Biochemical mechanisms underlying Collective Cell Migration in Cancer Metastasis}
\author[1]{Ushasi Roy \thanks{ushasiroy@iisc.ac.in}}
\author[2]{Tyler Collins} 
\author[1]{Mohit K. Jolly}
\author[2,3]{Parag Katira \thanks{pkatira@sdsu.edu}}
\affil[1]{Centre for BioSystems Science and Engineering, Indian Institute of Science, Bangalore, India}
\affil[2]{Department of Mechanical Engineering, San Diego State University, USA }
\affil[3]{Computational Science Research Centre, San Diego State University, USA }

\date{}

\begin{document}


\maketitle

\tableofcontents

\begin{abstract}

Multicellular collective migration is an ubiquitous strategy of cells to translocate spatially in diverse tissue environments to accomplish a wide variety of biological phenomena, viz. embryonic development, wound healing and tumor progression. Diverse cellular functions and behaviours, for instance cell protrusions, active contractions, cell–cell adhesion, biochemical signaling, remodeling of tissue micro-environment etc., play their own role concomitantly to have a single concerted consequence of multicellular migration. Thus unveiling the driving principles, both biochemical and biophysical, of the inherently complex process of collective cell migration is an insurmountable task. Mathematical and computational models, in tandem with experimental data, helps in shedding some light on it. Here we review different factors influencing Collective Cell Migration and then focus on different mathematical and computational models - discrete, hybrid and continuum - which helps in revealing different aspects of multicellular migration. Finally we discuss the applications of these modeling frameworks specific to cancer.
    
\end{abstract}

\section{Introduction - Collective Cell Migration}

Collective migration has been well-investigated in scenarios of embryonic developmental and wound healing, and has recently garnered attention in cancer metastasis as well, which has been largely thought of as driven by individual carcinoma cells undergoing an Epithelial- Mesenchymal Transition (EMT). In vivo observations about clusters of circulating tumor cells (CTCs) being the main drivers of metastasis \cite{aceto2014circulating} has driven much interest in investigating the biophysical and biochemical modes of formation of these clusters \cite{boareto2016notch} prior to dissemination from primary tumor, and their ability to traverse through capillary-sized vessels \cite{au2016clusters}. CTC clusters can contain cells of varying phenotypes – ranging from more epithelial or more mesenchymal – but hybrid epithelial/mesenchymal (E/M) cells seem to be better posited to form clusters \cite{bocci2019biophysical}. These clusters can not only contain cancer cells, but also various stromal ones and platelets \cite{lim2021circulating} which may aid in immune evasion through the journey. Both homotypic and heterotypic CTC clusters are highly metastatic and often correlate with worse patient outcomes across cancer types, as compared to the individual CTCs. Biochemically, these clusters have been interrogated at transcriptomic and methylation levels \cite{gkountela2019circulating}, while biophysical models have attempted to explain the size distribution of such CTC clusters \cite{bocci2019biophysical} as experimentally reported in cancer patients. These models have endorsed earlier observations of how intermediate cell-cell adhesion may maximize the chances of collective cell migration \cite{roy2021intermediate}. Further, in diverse contexts, a conceptual overlap of hybrid E/M phenotype(s) (moderate cell-cell adhesion levels) and collective cell migration (including the CTC clusters) is now being substantiated by identification of their underlying molecular basis \cite{vilchez2021decoding}.
\newline
\newline
In diverse contexts where collective migration has been probed further, questions on leader-follower traits have taken importance \cite{vilchez2021decoding}. It has been shown that leader and follower cells can often exchange their positions, revealing complex interplay at biochemical and biophysical levels within the cell population \cite{vishwakarma2020mechanobiology} – inter-cellular force transduction, energetic coordination etc. Thus, it becomes imperative to understand how various factors crosstalk to enable cells to behave collectively at multi-cell length scale in epithelial monolayers and/or sheet migration. Not only epithelial cells, but also mesenchymal cells can undergo collective migration \cite{theveneau2013collective}. For instance, in mesoderm – a mesenchymal tissue – cells migrate together with endodermal cells during gastrulation. Similarly, neural crest cells can migrate throughout the embryo by involving mesenchymal or hybrid E/M phenotypes \cite{theveneau2013collective}, involving contact inhibition of locomotion (CIL). Further, cells can often switch and back between collective and individual modes of migration too, including a switch to an amoeboid phenotype – cells that are deformable and soft, and often migrate without remodeling the extra-cellular matrix (ECM) by proteolysis \cite{wu2021plasticity}. Thus, cells undergoing collective-to-amoeboid transition (CAT) and vice versa amoeboid-to-collective transition (ACT) have been reported \cite{huang2015modeling}, besides the switch from mesenchymal to amoeboid and/or epithelial and vice versa: Mesenchymal-Amoeboid Transition (MAT), and Amoeboid-Mesenchymal Transition (AMT) \cite{talkenberger2017amoeboid}, as well as EMT and its reverse Mesenchymal-Epithelial Transition (EMT).
\newline
\newline
Together, given the complexity and plasticity of cell migration modes, decoding the multi-scale emergent dynamics of collective cell migration – characterized by multicellular groups migrating while retaining cell-cell junctions and a front-rear polarity – remains an active area of investigation in multiple biological systems, including cancer metastasis.

 \section{Factors influencing Collective Cell Migration}

At the individual level, there are two main mechanisms by which cells migrate on 2D substrates and through 3D environments \cite{lauffenburger1996cell, brabek2010role}. The first one is the mesenchymal mode of migration whose hallmarks are cell protrusions containing aligned actin-myosin fibers that bind and contract against the substrate \cite{bear2014directed}. The cell protrusions are transient and can be extended from the cell in different directions. The direction, frequency and length of these protrusions are governed by a variety of cellular and extra-cellular biochemical and mechanical factors.  Depending on the amount of force generated by the actin-myosin fibers within the protrusions, the strength and size of adhesions between the cell protrusions and the substrate and the mechanical properties of the substrate itself, the cell can pull itself along these protrusions. The second, commonly observed mode of migration is the amoeboid migration, where active decoupling of the actin-myosin cortex from the cell membrane causes the cell periphery to bleb. The actin polymerization and actin-myosin cortex recoupling with the cell-membrane within the bleb generates directional cortical tension pulling the cell body towards the bleb \cite{lim2013computational, charras2008life}. This motion also requires transient attachments between the cells and the substrate and can be equally fast as the mesenchymal mode of migration \cite{cox2016removal}. Cells have been shown to migrate using either mesenchymal or amoeboid or a combination of both these mechanisms \cite{tyson2014blebs}. They can also switch between these two mechanisms based on the mechanical and biochemical signals within the environment \cite{bravo2012directed,}. Beyond these two common mechanisms, there are a few other lesser-known mechanisms which have also been shown to drive individual cell migration such as osmotic pressure difference at the anterior and posterior regions of the cell, or asymmetric cell squeezing through undulating spaces \cite{lammermann2008rapid, petrie2014generation} . These mechanisms have been shown to be independent of cell-substrate adhesions, however, active forces within the cells are still known to play a role. Further details on mechanisms of individual cell migration and models describing them can be found in several recent reviews \cite{yamada2019mechanisms, conway2019cell, prahl2018modeling}. 
\newline
\newline
The most pertinent question with regards to collective cell migration that arises is how do individual cells that can migrate independently via the above-described mechanisms interact with other cells in close proximity to synchronize their speed and direction of migration to give rise to collective migration. To understand this question, we will focus on the various ways by which neighboring cells interact with each other –

\subsection{Direct Cell-Cell Mechanical Interactions} 
When two migrating cells collide with each other, they can either reverse direction/bounce away, go past each other, or begin to migrate together in the same direction \cite{scarpa2013novel, davis2012emergence, desai2013contact, milano2016regulators, camley2014polarity}. The precise outcome of any such cell-cell interaction is hard to predict but is shown to depend on cell-cell adhesion type, adhesion strength, cell migration speed, cell-substrate adhesion strength, cell actin-myosin contractility, cell polarity and the angle of approach between the cells among other factors \cite{singh2021rules}. A number of these factors themselves might be inter-related. For example, cell-cell adhesion is a function of the type of cell adhesion molecules present on the cell surface, the actin-myosin cortex tension, availability of other junction proteins as well as upstream intra-cellular signaling. On the other hand, cell polarity is dynamic and can change as a result of cell-cell collisions and interactions.
\newline
\newline
Mechanically, when two or more cells come into contact, they can form a number of stable or transient bonds depending on the cell type \cite{friedl2017collective, mayor2016the}. For example, epithelial cells generally form stable adhesions that permit the transmission of active contractile forces generated by the actomyosin cytoskeleton across the interacting cells. On the other hand, mesenchymal cells are known to form more transient bonds, that may only weakly transmit the cytoskeletal forces generated by one cell to another \cite{wong2014collective}. Partially transformed epithelial cells would then have a combination of stable and transient interactions with each other, allowing transmission of active cytoskeletal forces up to certain threshold values, and above which the cells can transit past each other and exchange neighbors or even migrate individually \cite{xi2019material}. 
\newline
\newline
Beyond cell-cell mechanical adhesions, a direct mechanical interaction between two cells can result in the repolarization of the cell. Migrating cells show a front-back polarity where the front end of the cell has faster actin polymerization, cell-substrate bond maturation, focal adhesion complex formation and increased recruitment of actin-myosin stress fibers, while the rear of the has increased dissolution of cell-substrate adhesions and increased actin depolymerization \cite{alexander1999self, bosgraaf2009navigation}. A direct mechanical interaction between two cells can lead to a change/reversal of polarity within these migrating cells causing them to migrate away from each other. This phenomena is known as Contact Inhibition of Locomotion (CIL) \cite{mayor2010keeping, abercrombie1979contact}. CIL in conjunction with transient cell-cell adhesion has been indicated in a variety of collective cell migration behaviors such as flocking, rotation of cellular clusters within confinement and even cells chasing or following each other \cite{stramer2017mechanisms, george2017connecting, lin2018dynamic}. 
\newline
\newline
When cells mechanically interact with each other, there are a few non-specific cell-interactions as well that can play important role in directing the dynamics of collective cell migration. For example, there can be a weak repulsion between cells due to the steric interactions between cell membranes and the exclusion of cellular volumes \cite{xi2019material}. Depending on the mechanical properties of the cell membrane, the actin-myosin cortex, the cell cytoskeleton and even the nucleus, these steric forces can lead to changes in cell shape and consequently changes in cell motility as well as proliferation. Additionally, non-specific interactions between membrane coating polymers such as the glycocalyx of interacting cells can dictate inter-cellular friction and alter the rates of cell migration within the collectives \cite{sabri2000glycocalyx, yao2007glycocalyx}. 

\subsection{Direct Cell-Cell Biochemical Interactions} Beyond direct mechanical interactions, the contact between two cells can trigger a cascade of biochemical signaling within the interacting cells. For example, active force transmission across cell-cell adhesion can result in triggering the $\beta$-catenin-Wnt signaling pathway which can trigger transcriptional changes driving stem-cell/cancer cell like behavior in epithelial cells \cite{BREMBECK200651, bienz2005beta}. Another example is the cell-cell interaction area dependent changes in Notch signaling which is critical in a variety of cell differentiation and boundary formation processes in tissues as well as tumor cell migration and invasion \cite{shaya2017cell}. 
\newline
\newline
The above-mentioned mechanical interactions between cells are also not independent of associated biochemical signaling. A key example of this is CIL, where the repolarization of interacting cells is associated with Eph/Ephrin interactions at cell-cell interfaces which trigger downstream RhoA activation and Rac inhibition \cite{batson2013regulation, astin2010competition}. Rac inhibition suppresses f-actin assembly and drive repolarization of the cell away from the cell-cell contact. Adhered cells can also communicate via Gap Junctions that transmit electrical signals as well as small molecules and effectors such IP3 and Ca+2 ions which act as secondary messengers for a variety of biochemical pathways that can promote of restrict collective cell migration \cite{haeger2015collective}. 

\subsection{Indirect Cell-Cell Interactions via the Environment} 
In the context of collective cell migration, the focus is primarily on direct cell-cell interactions and communication via mechanical and biochemical pathways. However, cells can also alter their immediate environment, and the sensing of these changes by neighboring cells can drive coordination and collective behavior between these cells. For example, contractile cells can strain the immediate extracellular matrix fibers \cite{sapir2017talking}. This strain can be sensed by other cells in the vicinity triggering strain sensitive migration and synchronization of contractile activity in these cells \cite{sopher2018nonlinear}. Alternatively, certain migrating cells can restructure the extracellular matrix via the action of matrix degrading enzymes or by depositing and crosslinking new matrix material. This active restructuring of the extracellular environment can create tracks or open spaces for other cells to follow, promoting collective migration and invasion of the extracellular environment \cite{kumar2016proteolytic, van2018mechanoreciprocity}.
\newline
\newline
Paracrine signals secreted by certain cells in the environment can also induce synchronization and collective cell migration of neighboring cells \cite{vilchez2021decoding}. An important example of this is the secretion of VEGF signals by chondrocytes that trigger collective cell migration of endothelial cells to for new vasculature \cite{apte2019vegf}. Cancer associated Fibroblasts also can release such paracrine signals such as cytokines and chemokines triggering collective invasion of cancer cells into the surrounding stroma \cite{bachelder2002vascular}. 

\subsection{Collective Cell Migration beyond Cell-Cell Interactions} 
\begin{figure}
    \centering
    \hspace*{-1.9in}
    \includegraphics[width=25cm]{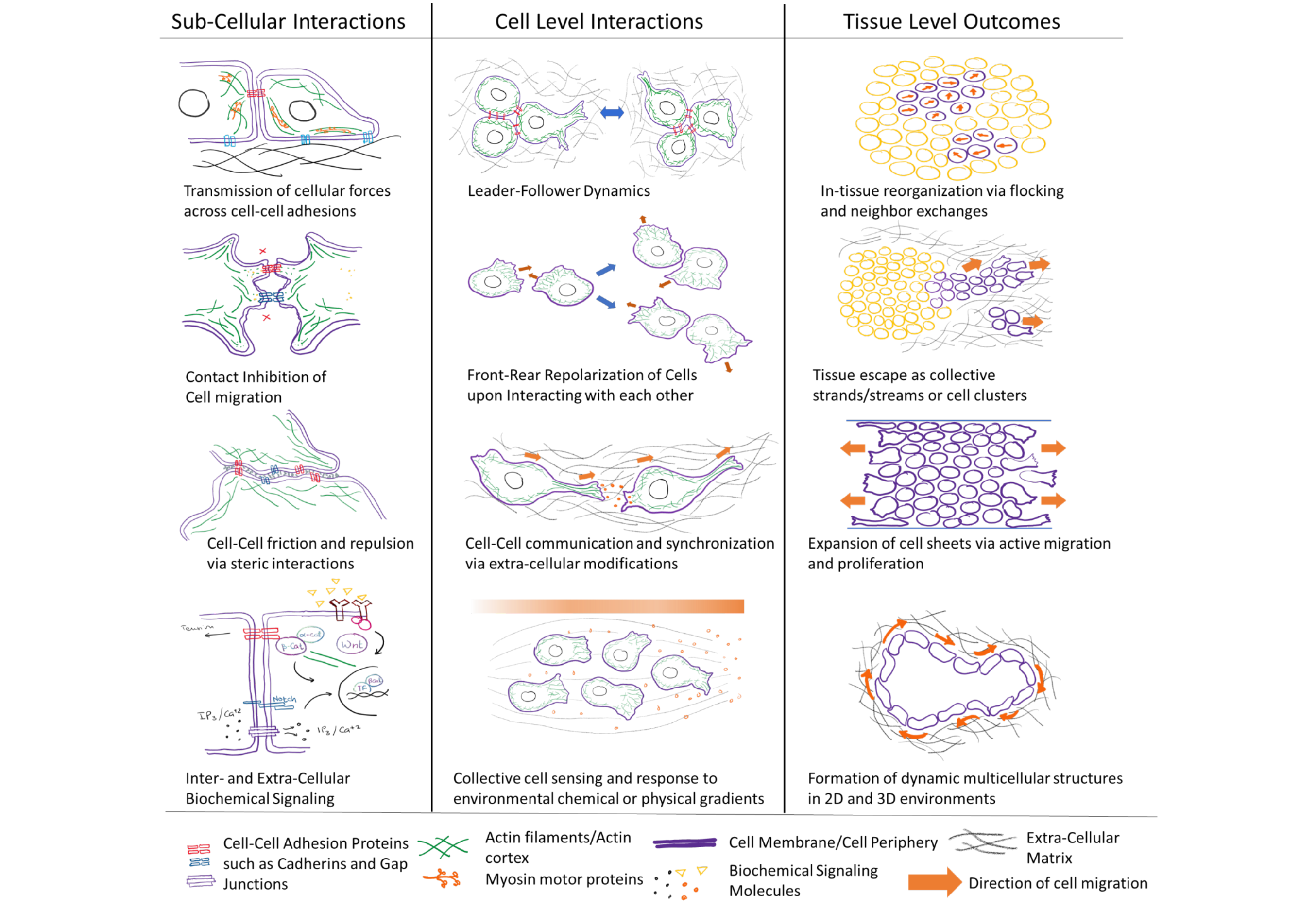}
    \caption{Multi-scale interactions and common outcomes of collective cell migration phenomena}
    \label{Figure1}
\end{figure}

While the common assumption is that synchronization of cell motility directions and speeds is an outcome of direct or indirect cell-cell interactions, it is also possible that the main drivers of collective motility are factors not affected by cell-cell interactions. One such scenario is the migration of cells under the influence of a chemical gradient \cite{roussos2011chemotaxis}. While each cell individually senses and moves towards or away from the chemical source, there does not have to be any cell-to-cell communication for the cells to move collectively. Similarly, cells migrating in confined spaces, along patterned substrates, along gradients of cell-substrate adhesion molecules or along gradients in substrate stiffness, all of these can cause collective cell migration independently of cell-cell interactions \cite{roca2013mechanical}. However, there are also scenarios where the presence of a cellular collective rather than individual, disconnected cells, promotes such directional migration along chemical and physical cues. For example, there is experimental evidence that collectives of cells can be better at recognizing and migrating along biochemical gradients than individual cells in a specific environment \cite{camley2016emergent}. Under which conditions individual cell behavior dominate vs collective cell interactions dominate the directional migration of cells is not completely understood. 
\newline
\newline
Ultimately, it is naïve to assume collective cell migration is an outcome of just one of the above interactions. In any given collective cell migration scenario, more than one of the above interactions can be influencing cell migration behaviors. Additionally, the above list of interactions is in no way exhaustive. We have just focused on some of the more commonly described phenomena known to drive collective cell migration. The combination of these cell-cell and cell-environment interactions occurring over a wide range of spatial and temporal scales results in the various collection cell migration phenomena observed in biological systems. These include – migration of small cell clusters, migration of large cell clusters, migration of cell strands (single-file), migration of cell streams (multiple cells wide), migration of cell sheets, rotation of cell clusters in confined environments, rotation of entire multicellular organoids, migration of cell along chemical gradients (chemotaxis), along stiffness gradients (durotaxis), along adhesion gradients (haptotaxis), along strain gradients (plithotaxis), herding via multiple leader cells, following a single or a small group of leader cells, flocking, jamming to unjamming transitions and many others. Figure \ref{Figure1} summarizes the multiscale nature of collective cell migration and highlights some of the key interactions and outcomes of cell-cell interactions and collective cell migration. The following section will discuss how biophysical and mathematical models can be employed to understand some of these processes.

\section{Discrete, Continuum and Hybrid Models}

The complexity of collective cell migration phenomena as described above, can be better understood with the help of computational models that allow the modeler to focus on a narrow set of interaction rules between cells and their environment at any given time and a limitless control over the parameter space. However, this means the models also have to make assumptions and simplifications regarding all the other interactions rules and properties not directly probed by the model system. Traditionally, computational complexity and costs have placed the largest constraints on the number of interactions incorporated within a model and the size of the parameter space. However, with advances in high-performance-computing, current models describing collective cell migration are highly integrated and incorporate a large variety of interactions and effects ranging from sub-cellular to tissue level in length scales and milliseconds to days in time scales. Models can be classified into discrete, continuum and hybrid models. “Continuum systems” and their canonical description can be mathematically depicted with nonlinear continuous ordinary differential equations (ODEs) or partial differential equations (PDEs) having multiple terms, each describing a biological process (viz. birth, death (apoptosis/necrosis), nutrient consumption etc.). But, in contrast, there is no well-defined way for describing interactions of discrete objects, particularly interacting and migrating discrete biological cells. In order to study the impact of the dynamics of an individual, multicellular cluster migration and cell-cell interactions on tissue behaviour, space- and time- dependent on- or off-lattice agent-based models (ABMs) needs to be developed. Sometimes we need to couple these two approaches and come up with a hybrid model to unleash the maximum knowledge about the relevant system with optimum computational effort. In the following sections we discuss some of the current computational models for collective cell migration.


\subsection{Isotropic Active Particles Model}
A standard minimal way of representation of interacting biological cells in a collective is by isotropic particles - all cells are of same circular/spherical shape in 2-/3-dimension and they interact in a similar fashion. The governing dynamics for the centre of mass of the overdamped (moving towards equilibrium) cell $i$ is given by the Newton's law with zero acceleration
\begin{equation}
    F_{\text{friction}}^i + F_{\text{active}}^i + F_{\text{cell-cell}}^i = 0
\end{equation}
where $ F_{\text{friction}}^i = - \gamma \frac{d \mathbf{x}_i (t)}{dt} $, $F_{\text{active}}^i = - \frac{\partial x_i}{\partial t} $ and $F_{\text{cell-cell}}^i =  -\frac{\partial}{\partial x_i} U(\mathbf{x}_1, ..., \mathbf{x}_N) $
where $U=\frac{1}{2}\sum_{i\ne j}|V_i-V_j|$. The first term $F_{\text{friction}}^i$ corresponds to the frictional drag of cells against substrate or extracellular matrix (corresponding to flow at low Reynold's number regime). This mathematical form expresses frictional force proportional to the velocity of a particular cell relative to a fixed background. It can also be modeled alternatively, which will depend on the relative velocities of cells in contact, e.g. $|v_i - v_j|$  \cite{palsson2000model, hoehme2010cell} or in a more complicated form based on dissipative particle dynamics \cite{basan2013alignment}. The second term represents the motility force, in the direction of the polarity of the cell. $U$ in the third term corresponds to a pair-wise interaction potential with V having long-range attraction and short-range repulsion between $i^{th}$ and $j^{th}$ cells. The effective long-range attraction represents adhesion of cells to their neighbors while short-range repulsion arises from an exclusion of the overlaps between different cells. However, cadherin mediated physical attraction is relatively short-range.

Herding, flocking, schooling, and swarming are different nomenclatures for the phenomena of interactive collective behavior, specific to animals, birds, fishes, and bacteria (or aggregate of any physical entities in general) respectively. Over the duration of the last twenty-five years, several self-propelled particle (SPP) models, which falls in the specific class of individual-based model (IBM)) have been introduced, studied and analysed, in one dimension \cite{czirok1999collective} and in higher dimensions \cite{couzin2002collective, cucker2007emergent, vicsek1995novel} with more biologically relevant interaction rules \cite{AOKI1982, couzin2002collective, gazi2004class, huth1992simulation, mogilner2003mutual, reynolds1987flocks}.

\subsection{Deformable Particle Model}


The next natural extension is to include cell-shape variability (in terms of small deviations from circularity/sphericity) in addition to its isotropic property. Cells were considered deformable ellipsoids of constant volume in models of Dictyostellium
\cite{camley2017physical}. Complex cell–cell interactions can be considered in these models. The authors \cite{palsson2000model, palsson2001three} have considered the interactions between the ellipsoidal cells as a function of the distance between cell surfaces - a natural generalization of the central forces applied in simple isotropic cell models. These models are valuable to study the coupling between shape and motility of cells. These models are most interesting when studying experiments that include both subconfluent and confluent layers of cells, rather than for purely confluent tissues which are better understood by Vertex/Voronoi models (described in Sec. \ref{Vertex Model}).

The net force experienced by each cell is given by

\begin{equation}
    F^i_{\text{Net}}=F^i_{\text{Active}}+\sum_{j\in N(i)} F^{ij}_{\text{Passive}}
\end{equation}

During the contraction phase of a cell, the force generated equals  $F^i_{\text{Active}}$. $F^{ij}_{\text{Passive}}$ is the passive force vector between cell $i$ and $j$. It has two origins, viz. compression - which is repulsive in nature, that arises from a cell's resistance to deformation, and adhesion - which is attractive, and the magnitude depends on their proximity, because it determines the number of adhesion molecules which can bind.

\subsection{Subcellular element models}
Biological cell is considered as the smallest unit in most models; which does not possess the capacity to capture detailed intra-cellular characteristics, e.g., acto-myosin contractions, cell morphology, cell polarization, cell size, cell division etc. To resolve this drawback, subcellular-element model does not consider each cell as the smallest unit, instead it has internal structures or sub-units interacting among themselves \cite{sandersius2008modeling, basan2013alignment, sandersius2011emergent, gardiner2015discrete}. In multicellular systems, this framework primarily computes the dynamics of large number of three-dimensional deformable cells. Each cell can be represented by a collection of elastically coupled elements, interacting with one another via short-range potentials. The dynamics of each element is updated
using over-damped Langevin dynamics. Mathematically, it is necessary to have two terms describing the intercellular $U_{inter}(r)$ and intracellular $U_{intra}(r)$ interaction potential of elements. Various models have tackled the scenario in different ways taking into account different aspects of interactions, for instance, by considering simple motility forces \cite{basan2013alignment}, or differential rates of breaking and re-forming of intra-cellular connections at the cell front/back \cite{sandersius2011emergent} or modeling the interior and the cell membrane separately \cite{gardiner2015discrete}.

The simplest subcellular-element model is comprised of two subcellular elements or units, representing a single cell. The two units stand for the front and back end of the cell \cite{zimmermann2016contact}. Using a two-subcellular-element model, the role of the supracellular actomyosin cable around the wound during healing of a wound can be successfully deciphered \cite{yang2018role}. The overdamped dynamics of each element is mathematically computed as follows

\begin{equation}
    \xi v= f_{SP} + f_{C} + f_{R/A}
\end{equation}
where $f_{SP}$ is the self-propulsion force which is subjected to contact inhibition of locomotion (CIL) \cite{zimmermann2016contact} and balances the intracellular contraction $f_{C}$ between the front and the rear unit. $f_{R/A}$ is the inter-particle force of different cells, which is repulsive in nature in short distances and attractive at long distances and vanishes further away. This term depicts the volume exclusion and intercellular adhesions.



\subsection{Active Network Models - Vertex/Voronoi Models \label{Vertex Model}}
Network models describe epithelial tissues as networks of polygonal cells. Vertex models play an important role in gaining deeper understanding of how forces generated inside cells affect the morphology of the cells and hence the shape and mechanics of epithelial sheets \cite{porta2019cell}. In the Vertex Model, a cell is characterized by a set of vertices at the intersection of three or more neighboring cells. The positions of the vertices further specify the cell interfaces and volumes.

Vertex models may be categorized into 2D and 3D apical vertex models, 2D lateral vertex models and 3D vertex models, based on the geometrical representation of the cells and how the forces act on the cells \cite{alt2017vertex}. In flat epithelium, forces are usually generated in parallel to the apical surfaces and Apical Vertex Models can be employed in understanding this phenomena, while 2D lateral vertex models are appropriate when the main forces act to deform the tissue in the plane of the cross-section. The tissue shape approximately remains the same under in-plane translation, and topological rearrangements do not play a role. 3D vertex models can be applied in a larger class of instances, viz., multicellular morphogenesis with undulation, tubulation, and branching \cite{okuda2018combining}, epithelial shape changes characterized by out-of-plane mechanics and three-dimensional effects, such as bending, cell extrusion, delamination, or invagination \cite{alt2017vertex}.

Contrastingly, the Voronoi Model is based on Voronoi construction \cite{honda1978description}, in which a cell is defined by its center and any point within the region of this cell is closer to this cell’s center than any other cell’s center. A Voronoi diagram is quite similar to the Wigner-Seitz-cell \cite{wigner1933constitution} description in solid state physics. The key difference between the Vertex and the Voronoi model is tracking the forces at the vertices versus energy tracking of whole cell based on shape and size. To investigate the collective behavior, a term for mechanical energy, having the cell’s area and perimeter, for each cell is included. Further insights on intercellular adhesion can be obtained from this energy. A self-propelled Voronoi model was developed \cite{bi2015density, bi2016motility} to demonstrate glassy dynamics and jamming transition from a solid-like state to a fluid-like state in a confluent biological tissue. The total energy $E$ of both the Vertex and Voronoi model systems can be expressed in the following (or similar) form in terms of the area and perimeter of each cell, positioned at $r$, is given by
\begin{equation}
E = \sum_{i=1}^N \Big[ K_A (A_{r_i} - A_0) + K_P (P_{r_i} - P_0) \Big]
\end{equation}
where $K_A$ and $K_P$ are the area and perimeter moduli, $A_{r_i}$ and $P_{r_i}$ are the cross-sectional area and perimeter of the $i^{th}$ cell whose center of mass is positioned at $r_i$, and which tends to relax to the preferred area and perimeter of $A_0$ and $P_0$ respectively \cite{porta2019cell}. The cell shape is defined by the Voronoi tessellation of all cell positions. This has provided a
good representation of epithelia of real biological systems, for instance the blastoderm of the red flour beetle \textit{Tribolium castaneum} and the fruit fly \textit{Drosophila melanogaster} \cite{vazquez2017mechanics, yang2017correlating}.

\subsubsection*{Key Insights: Transitions in Epithelia}

A generalized mechanical inference method \cite{yang2017correlating} have been developed by employing self-propelled Voronoi (SPV) model of epithelia, that connects mechanical stresses of cells to cell shape fluctuations and cell motility and deduce information regarding the rheology of the tissue. The interaction stress is generated from cell shape fluctuations due to actomyosin contractility and intercellular adhesion while cell motility
determines the swim stress that is generically present in all self-propelled systems. ‘Cellular Jamming’ is the collective arrest of cell movements and formation of a dense tissue. This phenomena is very similar to the process of ‘solidification’ or ‘rigidification’ in physical sciences - collective arrest of particle or molecular motion. Diverse physical mechanisms such as crowding, tension-driven rigidity, and reduction in fluctuations driven by biological features like cell-cell interaction, cell-substrate interaction, cell division, cell differentiation can lead to cellular jamming or crowding -  and one or more of these can simultaneously operate to have a 'phase transition' in a biological tissue \cite{yang2017correlating}.

\subsection{Cellular Automata}
Cellular Automata (CA) models can be viewed as ensembles of entities (cells) interacting with one another and the environment by phenomenological local rules, describing  biological processes. This is capable of modeling a huge range of biological examples ranging from bacteria, slime, amoeba, embryonic tisuues and tumors. Cellular Automata models are basically developed on a lattice which models interactions with other cells and the ECM.

Classical Lattice Gas based Cellular Automata (LGCA) Model, originally developed in 1973 by Hardy, Passis and Pomeau (HPP models) to model ergodicity-related problems to describe ideal gases and fluids, has been extended and widely applied to model self-driven biological cells \cite{alber2003cellular}. LGCA are relatively simpler CA models, in which entities (cells) select one from a discrete set of allowed rules. LGCA is capable of modeling a wide range of phenomena with different length- and time-scales \cite{hatzikirou2010prediction}.











\subsubsection{BioLGCA}
The BioLGCA model is a lattice-based agent-based cellular automata model class for a spatially extended system of interacting cells. BioLGCA models can be applied to homogeneous cell populations (cells having same phenotype and they don't change their behaviour). However, the BioLGCA can be expanded to model heterogeneous populations and environments, with cells dynamically regulating their adhesivities and/or interactions with a heterogeneous non-cellular environment. Several biological processes, for example, angiogenesis \cite{mente2011parameter}, bacterial rippling \cite{alber2004lattice}, active media \cite{syga2019lattice}, epidemiology \cite{fuks2001individual} and various aspects of tumor progression \cite{ilina2020cell, moreira2002cellular, reher2017cell, bottger2015emerging, hatzikirou2010prediction, tektonidis2011identification} have been studied by employing BioLGCA models. In BioLGCA, apart from biophysical laws for individual cell migration, update rules for cell-cell and cell-environment interactions can also be derived from experimental data of cell migration \cite{deutsch2021bio}, unlike update rules in classical LGCA which are mostly adhoc. This model combines mathematically rigorous analysis and computationally efficient simulations of collective cell migration, which has a scope to build rules based on experimental data. Thus it is capable of capturing complex multiscale behavior of collective cell migration. BioLGCA does not have cell-shape as a model feature, hence it is apt for modeling cellular behaviors at low and moderate cell densities, unlike epithelial tissue. BioLGCA minimises model artefacts by optimising their computational efficiency, their synchronicity and explicit velocity consideration, compared to other different categories of cellular automata models. The dynamics of the BioLGCA model, as described by the authors in \cite{deutsch2021bio}, comprises of Propagation $\mathcal{P}$, reorientation $\mathcal{O}$, phenotypic switch $\mathcal{S}$, and birth/death operators $\mathcal{R}$, all of which follows conservation laws maintained by operator dynamics. The modeling strategy is based on multiple biophysical interaction rules (Random Walk, Alignment, Attraction, Contact guidance, Chemotaxis, Haptotaxis) for individual based and/or collective migration of cells \cite{deutsch2021bio}. Interaction rules can be derived mathematically from Langevin equation of self-propelled particle and the steady-state distribution from its associated Fokker-Planck equation. A mean field analysis of the BioLGCA aggregation model may be performed to predict the formation of cluster patterns. The results show four distinct region in parameter space, viz. (i) Diffusive or gas-like phase - cells moving around freely as individual units, (ii) Collective motion (active nematic) - cells in collection move with an overall directionality, (iii) Aggregate phase - cells are in static clusters forming cellular patterns, and (iv) Jammed or glass-like - cells cannot move collectively or form patterns, all the cellular dynamics are frozen similar to a crystalline solid \cite{deutsch2021bio}.

\subsubsection{Cellular Potts Model}
A more sophisticated version of Cellular Automata (CA) is Cellular Potts Model (CPM) which considers individual cells as extended entities of variable shape \cite{alber2003cellular}.
The concept is borrowed from the q-state Potts Model in Statistical Physics. Anderson, Grest, Sahni and Srolovitz studied the cellular pattern coarsening in metallic grains in the early 1980s by employing q-state Potts model. CPM numerically captures the migration of multicellular clusters in two dimensions \cite{graner1992simulation}. The CPM can model realistic features of cells during migration, such as changes in cell-shape and -size, rearrangement of cells within a cluster, and the dynamic seggregation or reaggregation of subclusters. Diverse biological phenomena like chemotaxis, cell sorting, endothelial cell streaming, morphological development and tumor progression have been modeled using the CPM \cite{szabo2010collective, thuroff2019bridging, kabla2012collective}.


Each cell having the same polarization/orientation is represented by multiple lattice sites $x$ with the same integral values for their lattice labels $\sigma(x)>0$ in a discrete two-dimensional lattice. Lattice label $\sigma(x)=0$ represents the empty lattice sites corresponding to the extra-cellular matrix (ECM), which provides an environment through which the cells migrate. The energy of the whole system $E_{CPM}$ can be expressed as,

\begin{equation} \label{ECPM}
E_{CPM} = \sum_{\langle x,x' \rangle} J_{\sigma(x),\sigma(x')} + \sum_{i=1}^N \lambda_A(\delta A_i)^2.
\end{equation}

It has contributions from two factors: the first term denotes the adhesion while the second one is the area restriction term.
The adhesion energy term $J_{\sigma(x),\sigma(x')}$ is given by

\begin{eqnarray}
J_{\sigma(x),\sigma(x')} =
\begin{cases}
 0 & \;\;\;\;\;\, \sigma(x)\sigma(x) \; \ge 0 \; \textrm{within ECM or same cell},  \nonumber \\
 \alpha & \;\;\;\;\; \sigma(x)\sigma(x') = 0 \;  \textrm{cell-ECM contact}, \nonumber \\
\beta & \;\;\;\;\; \sigma(x)\sigma(x') > 0 \; \textrm{cell-cell contact}.
\end{cases}
\end{eqnarray}

$\alpha$ is the interaction strength of adhesion of any cell with its environment while cell-cell adhesiveness is characterized by $\beta$. A migrating cell undergoes fluctuations in size $\delta A_i$ around its equilibrium area $A_0$ ($\delta A_i \equiv A_i(t) - A_0$). The second term, i.e., the area restriction term in Eq. \eqref{ECPM} controls a migrating cell from growing or shrinking to unphysical sizes, as well as branching or stretching into intangible shapes.
A perimeter restriction term, in addition to the area restriction term, has also been included in many works \cite{glazier1993simulation, graner1992simulation, varennes2016collective, camley2017physical}. But this might be omitted for simplicity since sufficiently large $\alpha$ and $\beta$ constrain perimeter by cell-ECM or cell-cell contact.
Two cells in contact with each other will have an adhesion energy cost of $\beta$, while if the two edges of these cells are exposed to the ECM they will have an energy cost of $2\alpha$. Thus we have two regimes: adhesion energy satisfying the condition $\beta < 2\alpha$ would promote cell scattering, while $\beta > 2\alpha$ will promote clustering. In between, there lies the optimum regime of $\beta \approx 2\alpha$ which corresponds to the transitional state between a fully connected cluster and multiple disconnected clusters he optimal migration velocity corresponds. in this regime, the multicellular cluster attains the maximum effective migration velocity \cite{roy2021intermediate}.










\subsection{Phase-field model}
Another approach developed by many research groups \cite{camley2017physical, camley2014polarity, nonomura2012study, lober2015collisions, palmieri2015multiple} is to model shape variability of cells during collective motion with space- and time-dependent phase fields $\phi (\mathbf{r}, t)$. This model can simulate relatively small ($<100$) collections of cells, but can capture the scenario of simultaneous couplings among the biochemistry, cell shape, and mechanics.
The minimization of the Hamiltonian $\mathcal{H}$, with added active terms representing the motility of the cell, gives rise to the equations of motion for the phase field, which can be expressed in the following form
\begin{equation} \label{EOM_PhaseField}
    \frac{\partial}{\partial t} \phi (\mathbf{r},t) + \mathbf{v}_{active}.\nabla \phi= - \frac{1}{\zeta \epsilon} \frac{\delta \mathcal{H}}{\delta t}
\end{equation}
Here $\mathbf{v}_{active}$ is the velocity due to active driving at the boundary. The driven active velocity may be constant and in a direction parallel to the cell’s polarity, or to be normally outward, with a magnitude set by biochemical polarity within the cell. $\zeta$ is the friction coefficient and $\epsilon$ width of the phase field.
The Hamiltonian $\mathcal{H}$ in the above equation \ref{EOM_PhaseField} has contributions from two terms: $\mathcal{H} = H_{single} + H_{cell-cell}$. The first term for the single cell energies is given by the Canham-Helfrich energies (widely applied for the deformed fluid membrane) \cite{brown2008elastic} as follows:
\begin{eqnarray}
    H_{single} & = & \gamma \times \text{[cell perimeter]} + \kappa \times \text{[curvature integrated over membrane]} \nonumber \\
    & = & \gamma \int d^2 \mathbf{r} \bigg[ \frac{\epsilon}{2} |\nabla \phi^2| + \frac{G(\phi)}{\epsilon} \bigg] + \kappa \int d^2 \mathbf{r} \frac{1}{2\epsilon} \bigg[ \epsilon \nabla^2 \phi - \frac{G'(\phi)}{\epsilon} \bigg]^2
\end{eqnarray}
where $\epsilon$ is the free parameter characterizing the width of the interface of the the phase-field, $\gamma$ is the tension.
And the second term corresponds to the bending energy of the membrane computed by the integral over the mean curvature of the membrane.
\begin{equation}
H_{cell-cell} = \sum_{i \ne j}\int d^2 \mathbf{r} \frac{g}{2} \phi^{(i)}(\mathbf{r},t)\phi^{(j)}(\mathbf{r},t) - \frac{\sigma \epsilon^3}{4} | \phi^{(i)}(\mathbf{r},t) | ^2  | \phi^{(j)}(\mathbf{r},t) | ^2
\end{equation}
where $\kappa$ is the bending modulus of the membrane.
Camley \textit{et al.} investigated the effect of various cell polarity mechanisms on the rotation motion of a pair of mammalian cells, including contact inhibition of locomotion, alignment of position or velocity with neighboring cells \cite{camley2014polarity} that can be experimentally observed \cite{segerer2015emergence}. They showed that the persistent rotational motion is promoted by the velocity alignment robustly. Lober \textit{et al.} came up with an alternative phase-field model to simulate hundreds of cells \cite{lober2015collisions}.
Phase field models have the advantage that they can model higher-order nonlinear terms like the bending energy of the membrane and can be readily integrated with reaction–diffusion mechanisms. However, they have a high computational cost, since the dynamics of each cell would follow a different partial differential equation which needs to be numerically solved.




\subsection{Finite Element Immersed Boundary Models}
This biomechanical model for a single fully deformable cell accounts for the interactions/couplings between the fluid flow of the viscous incompressible cytoplasm and the structural deformations of the elastic cell \cite{rejniak2007immersed}. It is postulated that the fluid flow is governed by the Navier-Stokes equation, given by \cite{batchelor2000introduction}
\begin{eqnarray}
    \rho \nabla . \mathbf{u} &=& s \nonumber \\
    \rho \bigg( \frac{\partial \mathbf{u}}{\partial t} + (\mathbf{u} . \nabla ) \mathbf{u} \bigg) &=& - \nabla \mathbf{p} + \mu \Delta u + \frac{\mu}{3\rho} \Delta s + \mathbf{f} 
\end{eqnarray}
Here, $\rho$, $u$, $s$ denote the constant fluid density, fluid velocity, fluid source distribution respectively and $\mu$ is the constant fluid viscosity, $p$ the fluid pressure, and $\mathbf{f}$ the external force density.
This is a description of the balance of momentum and the mass in a viscous incompressible fluid with distributed
sources. In this model, the fluid is assumed to be incompressible except at the sources which are representative of positions indicating cell-growth. Hence, on the whole fluid domain $\Omega$ except at the isolated point sources, the local fluid expansion rate $\nabla . u$ and the source distribution $s$ are identically equal to zero. The force density $f$ defined by the forces $F(l,t)$ at the elastic immersed boundaries of all cells $X(l,t)$ can be expressed in the following way
\begin{equation}
\mathbf{f} = \int_{\Gamma} F(l,t) - \delta(x-X(l,t)) dl
\end{equation}
The $\delta$ is the Dirac delta function, $\Gamma$ represents the finite collection of immersed boundary of all cells (the external force $\mathbf{f}$ vanishes away from $\Gamma$). The elastic cell membranes can be defined by the curvilinear coordinates $X(l,t)$ where $l$ denotes the position along the cell boundaries. The boundary forces $F(l,t)$ has a contribution from three different types of forces, viz., the adjacent forces $F_{adj}$, inter-cellular adhesion forces $F_{adh}$, and the contractile forces and are determined by the boundary configuration, the assumed elastic properties of the cell membranes and the undergoing cell processes.

This model has been applied to track morphology of the cell membrane during cytokinesis \cite{rejniak2007immersed}, a single axisymmetric cell growth and division \cite{li2012immersed}, blood flow and shear stresses in cerebral vessels and aneurysms \cite{mikhal2013development}.

\subsection{Hydrodynamic Models}

During embryonic development of an organism, mechanical stresses in a biological tissue develops due to cell division, cell apoptosis and other factors to maintain tissue homeostasis, which play a significant role in tissue growth and hence final tissue morphology \cite{delarue2014stress}. The hydrodynamic models systematically studies the effects of fields on tissue dynamics. The dynamics of a thick polar epithelial tissue subjected to the action of both an electric filed and a flow field in a planar geometry is theoretically studied \cite{sarkar2019field}. A generalized continuum hydrodynamic description is developed to describe the tissue as a two component fluid system - the cells and the interstitial fluid. A biological tissue in a continuum limit may be characterized by a cell number density $n$ which obeys the following equation to balance the cell number
\begin{equation}
   \partial_t n + \partial_{\alpha} (n v_{\alpha}) = (k_d - k_a) n
\end{equation}
where $v_{\alpha}$, $k_d$ and $k_a$ are the velocity, division rate and apoptosis rate for a characteristic cell in the tissue \cite{delarue2014stress}.
A "polar" and planar thick epithelial tissue, permeated by interstitial fluid flows in the presence of an electric field with cells able to generate electric currents and fluid flow, will follow conservation laws for volume, charge and momentum, symmetry considerations, and cell number balance equations \cite{sarkar2019field}. This predicts that the domain of stability of the epithelial tissue of finite thickness is rather small. Tissue proliferation or tissue collapse can be caused due to simple dc electric current or a fluid flow, without any requirement for genetic mutation.

\subsection{Simulation Platforms}
\begin{figure}
    \centering
    \vspace*{-1.4in}
    \hspace*{-1in}
    \includegraphics[width=20cm]{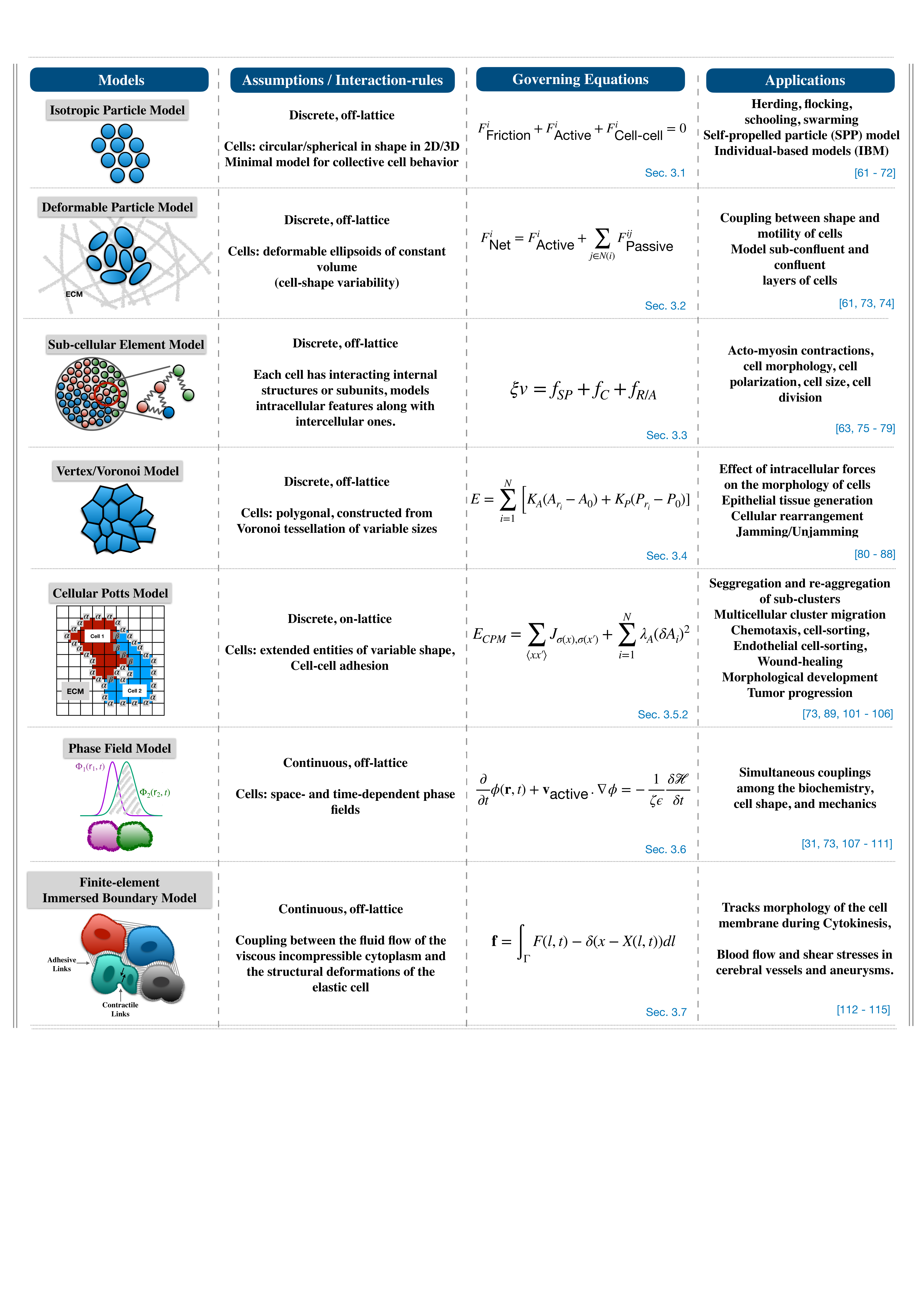}
    \vspace*{-2.4in}
    \caption{\textbf{Various Mathematical and Computational Models of Collective Cell Migration:} The key features of each model is highlighted along with the governing equations and applications.}
    \label{Figure2}
\end{figure}

Recently numerous softwares have been developed to capture \textit{in-silico} this inherently complex collective behavior of cells. Efficient computational agent-based multiscale models are developed to understand the underlying complexities at different level.
\newline
\newline
BioFVM is an efficient diffusive transport solver, written in C++ with parallelization in OpenMP, for solving systems of Partial Differential Equations (PDEs) in 3D for release, uptake, decay and diffusion of multiple substrates in multicellular systems \cite{ghaffarizadeh2016biofvm}.
PhysiCell - physics-based multicellular simulator — an open source agent-based simulator - is an ideal \textit{in-silico} "virtual laboratory" that is capable of modeling many mechanically and biochemically interacting cells in dynamic biochemical tissue-microenvironments \cite{ghaffarizadeh2018physicell}.
Markovian Boolean Stochastic Simulator (MaBoSS) is a C++ software for the stochastic simulation continuous/discrete time Markov processes by employing Monte-Carlo kinetic algorithm (or Gillespie algorithm), applied on a Boolean network, to compute global and semi-global characterizations of the whole system. A Python interface for the MaBoSS software, called pyMaBoSS, is also available \cite{stoll2017maboss}.
PhysiBoSS is an open source software, a multi-scale agent-based modelling framework, which integrates multicellular behaviour using agent-based modelling (PhysiCell) and intracellular signalling using Boolean modelling (MaBoSS) \cite{letort2019physiboss}.
CompuCell3D is another open source multi-scale multi-cellular computational modeling environment based on Cellular Potts Model, which can handle wide variety of problems including angiogenesis, bacterial colonies, cancer, embryogenesis, evolution, the immune system, tissue engineering, toxicology and even non-cellular soft materials \cite{swat2012multi}.
\newline
\newline
Thus, this section gives a detailed overview of most of the important mathematical and computational models to study various aspects of collective cell migration. Figure \ref{Figure2} illustrates different models of collective cell migration through simple schematics, highlighting their assumptions/interaction-rules, governing equations and applications.

\section{Biochemical Models}
Most efforts towards biochemical understanding of collective cell migration has focused on intracellular networks related to EMT, particularly those connected to hybrid E/M phenotypes. EMT has been reported in various contexts: embryogenesis (type I EMT), wound healing and tissue homeostasis (type II EMT), and cancer metastasis and fibrosis (type III EMT) \cite{kalluri2009basics}. It has been largely thought of as “all-or-none” binary process, but recent observations – both computational and experimental – have unraveled that cells can acquire stably one or more hybrid E/M phenotype(s) with distinct biochemical and/or biophysical signatures \cite{subbalakshmi2022biophysical}. Here, we discuss different modeling strategies that have been adopted to elucidate EMT dynamics.

\subsection{ODE based models}

Mathematical modeling of underlying biological networks has been instrumental in driving the appreciation of EMT from a binary process to a multi-step one. Both the initial models on EMT signaling were ODE-based, and modeled a set of experimentally identified interactions through coupled ODEs \cite{tian2013coupled, lu2013microrna}. Despite slightly different formalisms and parameter values used, they both predicted that cells can stably acquire a hybrid E/M phenotype. Both the models captured the dynamics of coupled feedback loops among the families of EMT and MET regulators: EMT-driving transcription factors ZEB and SNAIL, and MET-drivers microRNAs miR-200 and miR-34. Other ODE models that have expanded these networks to incorporate additional nodes have shown that more than one hybrid E/M states may exist, and that some molecules can act as ‘phenotypic stability factors’ (PSFs) for hybrid E/M phenotype(s) such as GRHL2, OVOL1/2, NRF2, NFATc \cite{subbalakshmi2020nfatc, jolly2016stability}. Interestingly, knockdown of these molecules – one at a time – in H1975 lung cancer cells (that display hybrid E/M phenotype stably \textit{in vitro} for multiple passages) can push them to a more mesenchymal phenotype, as detected based on morphological and molecular changes. Thus, these PSFs can be thought of as ‘molecular brakes’ on EMT, and can increase the residence times of cells in hybrid E/M phenotype(s) \cite{lu2013microrna}.

More recent ODE based models for EMT have incorporated additional contextualization, for instance, hypoxia-driven EMT such as HIF1$\alpha$ \cite{wang2021regulation} which includes SNAIL, TWIST and miR-210, and suggest that number of positive feedback loops in a regulatory network determines the number of steady states related to EMT seen. This observation is reminiscence of analysis of many EMT networks across a wide range of parameter values, showing that the number of positive feedback loops correlates strongly with the propensity of acquiring multistability \cite{hari2020identifying}.

\subsection{Boolean models}

As the network size grows, it becomes increasingly difficult to estimate the different kinetic parameters needed to simulate an ODE model. Thus, often, a parameter-free approach such as discrete/Boolean/logical modeling is used to decode the qualitative emergent network dynamics. In this formalism, each node in the network can be either ON (1) or OFF (0), and the levels of each node at a given time can be decided by other nodes that can either activate or inhibit it. For instance, if A inhibits B, then B (t+1) = OFF when A (t) = ON. This framework facilitates investigating the dynamics of larger networks, as has been in the case of EMT models too, and can identify different “attractors” a system can eventually converge to. Boolean models of EMT networks have also indicated co-existence of many hybrid E/M phenotypes, each defined by a set of specific nodes that are ON and others that are OFF \cite{steinway2015combinatorial, font2018topography, silveira2020systems}.

\section{Collective Cell Migration in Cancer}


Cells of a primary tumor undergo Epithelial to Mesenchymal Transition (EMT), become motile and invasive, intravasate through blood stream. At a distant secondary site, the cells undergo extravasation forming micrometastases, which is known as Mesenchymal to Epithelial Transition (MET). This property of cells to switch phenotypes is well-known as Epithelial Mesenchymal Plasticity.
\newline
\newline
Over the last two decades, Epithelial-Mesenchymal Transition (EMT) has been recognised as an essential, highly-regulated, and mostly-employed mechanism in normal embryological development and in migration of epithelial-derived cells in morphogenesis. Of late, \textit{in-vitro} and \textit{in-vivo} studies in tandem with that in \textit{in-silico} studies have suggested that EMT may also play an important role in cancer metastasis in various cancer systems. Recently, the production of fibroblastoid cells from epithelial precursors in fibrotic diseases has been recognised as bearing the hallmarks of EMT. The biological and pharmacological importance of EMT in cancer has now achieved much attention and recognition. Problems in multicellular systems can only be understood if we have a wholistic view of its overall biological activities, viz., interaction, motion, growth, division, and death and as a result of all these, multifarious tissue-scale dynamics evolve.
\newline
\newline
In cancer metastasis, EMT at the primary site causes sedentary epithelial cells to change their phenotype to become less confluent and more migratory (akin to a solid $\to$ fluid or jamming $\to$ unjamming transition (UJT)). During UJT, cells lose apico-basal polarity and epithelial markers, while they concurrently gain front-back polarity and mesenchymal markers. And during MET at secondary site, epithelial cells change their phenotype to become more confluent and loses motile behavior (akin to a fluid $\to$ solid or unjamming $\to$ jamming transition). In a mixture of both epithelial and mesenchymal cells, when only a fraction of the cells have transitioned to the second phenotype, the tissue achieves a frustrated jamming state \cite{castro2016clustering}. But recently, subtle differences between EMT and UJT have been pointed out in primary airway epithelial cells \cite{mitchel2020primary}. The claim is that, the collective epithelial migration can take place through UJT with or without EMT or partial EMT.
\newline
\newline
In experimental models for cancer metastases, it has been observed that in lung cancer \cite{liotta1974quantitative, kusters2007micronodular}, tumor cells survive in clusters in the blood stream and generate lung metastases, and in inflammatory breast cancer associated with lymphatic metastasis, multicellular strands travel through the lymphatic vessels \cite{kusters2007micronodular}. A balance between EMT/UJT, collective cell migration either as clusters of strands and MET/JT at specific invasion sites might hold the key to accurately predict disease outcomes in many cancers. Additionally, biochemical and mechanical factors triggering these collective cell phenomena are of extreme interest as potential diagnostic markers or therapeutic targets. 
\newline
\newline
In multicellular systems, Least microEnvironmental Uncertainty Principle (LEUP) \cite{barua2020least, pujar2021lattice} might be employed to understand the phenotypic decision-making of a cell, based on the phenotype of the neighboring cells. During cancer metastasis, cells travel as individual units and as multicellular clusters of circulating tumor cells (CTCs) through the blood stream to a distant secondary organ. While in transit, the cells are in a dynamically evolving micro-environment in which it changes its phenotype multiple times. In future, in the combined framework of LEUP and mechanistic models of collective cell migration, the phenomenological properties such as entropy, cell migration velocity, phase separation and non-equilibrium structure formation could be mapped to other phenotypic markers, thus bridging the gap between computational models, \textit{in vitro}/\textit{in vivo} experiments and clinical data.







\section{Conclusions and Future Directions}

Individual biochemical and biophysical models have been instrumental in decoding the mechanisms of collective cell migration, with important implications in unraveling metastasis. An integrated unique framework that combines the essential features of both the biophysical and biochemical aspects would help us understand the phenomena of cancer metastasis in a more concrete manner \cite{shatkin2020computational, deng2021theoretical}. Various aspects of mechanochemical coupling have begun to be unearthed in various contexts of collective cell migration \cite{hino2020erk, bui2019mechanochemical, murad2019mechanochemical}, and the next class of mathematical models shall benefit from incorporating those latest details \cite{hirway2021multicellular, boocock2021theory}.
\newline
\newline
One challenge that needs to be overcome is connecting diverse length and time scales: sub-cellular (intra-cellular components), cellular, supra-cellular/tissue-scale, and eventually multi-organ level (during cancer metastasis) and organism level (during embryonic development). Another aspect that needs attention is the granularity with which we model this phenomenon. Cells migrating collectively cannot be simply thought of as isotropic active particles following the exactly same generic principles. While this approximation can be a good minimal model to start with, various other factors need to be incorporated for biological realism: short-range and long-range communication via physical contact and/or diffusing biochemical molecules \cite{labernadie2017mechanically, vandervorst2019wnt, wyckoff2004paracrine}, change in microenvironment due to spatiotemporal variations in signaling molecules \cite{bocci2019toward, carmona2013emergence} etc. Thus, within a mean field theory framework, a stochastic term that encapsulates many of these differences to represent cell-to-cell variability would be an important consideration. A multi-scale agent-based framework can offer a good compromise in terms of granularity and usefulness of the model.
\newline
\newline
A multi-scale model often incorporates many timescales set by rate of biochemical reactions and those set by diffusivity of various sensory as well as cell-cell communication molecules \cite{yaron2014juxtacrine}. Models for tissue-level patterning that encapsulates intra-cellular biochemical signaling with inter-cellular communication have been well-investigated, but most of this this work has been in static scenarios \cite{bocci2020understanding}. Thus, coupling these frameworks with cell motility (through, say, vertex models) at individual and cohort levels, as seen during morphogenesis, can be a first step \cite{bajpai2021role} toward an integrated understanding of the multi-scale dynamics of collective cell migration.
\newline
\newline
Given the recent advancements in collecting longitudinal data at mechanochemical levels, an active crosstalk between mathematical models and experimental data can accelerate our progress in decoding the underlying principles of collective cell migration, including during the process of metastasis, and interfering with it to have potentially better patient outcomes. Many open questions remain: a) what type of model framework can best describe collective cell motility? b) why a multicellular cluster senses the environment better than its individual component, and is this reason a contributing factor to higher metastatic potential of clusters? c) what are the origins of cell-cell communication (autocrine, paracrine, juxtacrine signaling), diffusion of sensory molecules in extracellular environment? d) how does cell-cell adhesion and cell-ECM adhesion interplay in enabling collective cell migration with varying efficiency?, and e) what optimal cluster sizes are observed upon collective migration of cells during metastasis, and how can those size distributions be shaped by the underlying biochemistry?

\printbibliography

\end{document}